\documentclass{article}

\usepackage[preprint]{neurips_2026}

\usepackage[utf8]{inputenc} 
\usepackage[T1]{fontenc}    
\usepackage{hyperref}       
\usepackage{url}            
\usepackage{booktabs}       
\usepackage{amsfonts}       
\usepackage{nicefrac}       
\usepackage{microtype}      
\usepackage{xcolor}         
\usepackage{colortbl}
\usepackage{tcolorbox}
\usepackage{color}
\usepackage{tabularx}
\usepackage{multirow}
\usepackage{todonotes}
\usepackage{comment}
\usepackage{algorithmic}
\usepackage{graphicx}
\usepackage{textcomp}
\usepackage{tikz}
\usepackage{soul}
\usepackage[normalem]{ulem}
\usepackage{amsthm}
\tcbuselibrary{skins}         
\tcbuselibrary{breakable}     
\tcbuselibrary{theorems}

\definecolor{datarow}{RGB}{225, 236, 250}        
\definecolor{trainingrow}{RGB}{234, 225, 245}    
\definecolor{inferencerow}{RGB}{225, 245, 230}   
\definecolor{crossrow}{RGB}{235, 235, 235}       

\usetikzlibrary{positioning}

\newtcbox{\artifact}{on line, boxrule=0.4pt, colback=yellow!5, colframe=black!40,
  fontupper=\small\sffamily, boxsep=1pt, left=2pt, right=2pt, top=0.5pt, bottom=0.5pt}

\newtcbox{\transformation}{on line, boxrule=0.4pt, colback=blue!5, colframe=black!40,
  fontupper=\small\sffamily, boxsep=1pt, left=2pt, right=2pt, top=0.5pt, bottom=0.5pt}

\newtcolorbox[auto counter]{definitionbox}[1][]{
  colback=black!5!white,
  colframe=black!60!white,
  fonttitle=\bfseries,
  title=Definition~\thetcbcounter,
  breakable,
  enhanced,
  sharp corners,
  boxrule=0.8pt,
  #1
}

\title{The Grand Software Supply Chain of AI Systems}

\author{
  Carmine Cesarano, Martin Monperrus \\
  KTH Royal Institute of Technology\\
  Stockholm, Sweden\\
  \texttt{\{cesarano,monperrus\}@kth.se} \\
}

\begin{document}

\maketitle

\begin{abstract}
AI systems rest on software with low integrity mechanisms, leaving AI systems exposed across every stage from data acquisition to final inference. 
This paper makes the AI supply chain a first-class object of analysis, decomposing it across four architectural layers: data acquisition, model training, model inference, and a cross-cutting substrate. 
Within these layers, we identify four structural gaps that traditional supply chain mechanisms do not address: verifiability, versioning, observability, and traceability.
Current AI systems fall short on all of them:
they carry undeclared behavioral couplings that no resolver enforces;
they cannot be reverted back to known working assemblies;
they degrade silently rather than surfacing breaking changes; 
and their lineage can hardly be approximated. 
To illustrate the scale of the software supply chain of AI, we measure a reference stack of 48 production-grade open-source projects, which declares 4,664 direct dependencies, resolves to 11,508 transitive packages, and totals roughly 392M lines of code.
\end{abstract}

\section{Introduction}
\label{sec:intro}

The integrity mechanisms that secure traditional software, including reproducible builds, semantic versioning, signed attestations, do not transfer to AI systems, yet no framework exists to characterize what AI's supply chain even encompasses. The consequences are already visible: in recent years, outages and security incidents have accumulated across every layer of the AI stack.
In February 2024, researchers disclosed that roughly one hundred models hosted on Hugging Face, the de facto distribution hub for open-weight AI, carried malicious payloads that execute arbitrary code on any machine that loaded them~\cite{jfrog2024huggingface}. 
Two months earlier, the Stanford Internet Observatory had identified 1{,}008 instances of child sexual abuse material inside LAION-5B, a five-billion-sample dataset underlying Stable Diffusion and a long tail of derived image generators whose full extent nobody could enumerate~\cite{thiel2023identifying}. Early in 2026, a compromised dependency in LiteLLM's CI/CD pipeline allowed attackers to publish a backdoored release that exfiltrated cloud credentials, SSH keys, and Kubernetes secrets~\cite{wiz2026litellm}. And throughout this period, operators of applications built on frontier model APIs have reported concerning behavioral changes: on one widely cited benchmark, GPT-4's accuracy at identifying prime numbers fell from 84\% to 51\% over three months, with no version selector to roll back to and no publicly available changelog~\cite{chen2024chatgpt}.

These incidents are not four unrelated failures. They are four symptoms of the same underlying condition: the software supply chain of modern AI systems is large, heterogeneous, and not managed. A single production inference call results from the execution of data acquisition pipelines, training orchestration frameworks, model registries, prompt templates, guardrail filters, quantization toolchains, and serving stacks, each a distinct software project with its own dependency tree. 

This paper argues for making the software supply chain of AI systems a first-class object of analysis. We decompose the AI software supply chain into four architectural layers (data acquisition, model training, model inference, and cross-cutting substrate) and three abstractions (artifacts, transformations, and infrastructure). To make the scale of the problem concrete, we define and measure a reference AI stack of 48 typical open-source projects whose production use is independently documented. The reference AI stack declares 4{,}664 direct dependencies and resolves to 11{,}508 transitive packages, totaling roughly 392M lines of code. This is the scale of the AI software supply chain.

Next, we identify four structural gaps that the current supply chain fails to close. The \textbf{Verifiability} property fails today because trained artifacts cannot be reproduced bit-for-bit, so hash verification reveals nothing about whether a checkpoint was produced by the procedure its attestation describes. The \textbf{Versioning} property fails because compound AI systems carry undeclared behavioral couplings (adapter-to-base, index-to-embedder, prompt-to-output-format) that no resolver tracks and enforces, thus first-party API providers can rewrite the served artifact unilaterally. The \textbf{Observability} property fails because AI systems degrade silently rather than crashing, and current telemetry, focused on latency and error rates, fails to attribute a behavioral shift to its root cause. The \textbf{Traceability} property fails because lineage graphs in AI are multi-parent, non-deterministic, and cross-organizational, none of which existing attestation frameworks (SLSA, in-toto) are able to represent.
These four gaps define the research agenda this paper addresses.

To sum up, our contributions are:
\begin{itemize}
    \item the first-ever characterization of the AI software supply chain, which demonstrates its scale, depth and complexity;
    \item the definition of four essential integrity properties that current AI systems fail to meet, defining a coherent research agenda;
    \item an empirical measurement of a 48-project reference AI stack resolving to 11{,}508 transitive packages and roughly 392M lines of code, concretely establishing the scale of the problem, with an associated web-based visualization at \url{https://assert-kth.github.io/ai-supply-chain/}.
\end{itemize}

The remainder of the paper is organized as follows. Section~\ref{sec:ai_ssc} introduces the layer-and-abstraction framework. Sections~\ref{sec:data}--\ref{sec:cross-cutting} examine each architectural layer. Section~\ref{sec:challenges} formalizes the four gaps. Section~\ref{sec:scale} reports the empirical measurement of the reference stack. Section~\ref{sec:related} analyzes related work. Section~\ref{sec:conclusion} concludes the paper.

\section{The Software Supply Chain of AI Systems}
\label{sec:ai_ssc}

Modern AI systems combine myriad software components.
Those components are produced independently by multiple providers and integrated on shared infrastructure~\cite{compound-ai-blog}. 
We call the collection of these components the software supply chain of AI.

\begin{definitionbox}
The \textbf{software supply chain of an AI system} is the collection of software components that produce, transform, store, and serve the artifacts of that system, from the initial training data collection to the final inference run in production.
\end{definitionbox}

We decompose this software supply chain into four \emph{architectural layers}, each addressing a distinct concern in the construction and operation of a compound AI system. Three layers correspond to the principal concerns through which an AI system comes into being and is delivered: \emph{data acquisition} (Section~\ref{sec:data}) comprises the software that collects and curates datasets; \emph{model training} (Section~\ref{sec:training}) comprises the software that derives model checkpoints from datasets; \emph{model integration and inference} (Section~\ref{sec:integration}) comprises the software that composes checkpoints with prompts, retrieval pipelines, tools, and guardrails into deployable systems that serve end users. The fourth layer, \emph{cross-cutting infrastructure} (Section~\ref{sec:cross-cutting}), is not a stage-specific concern but a shared software substrate whose components appear as dependencies within the other layers. The three stage layers are not traversed linearly: deployed systems emit observations that feed back into data curation, and new data or newly observed failure modes trigger additional rounds of training, so components circulate continuously across layers.

We describe this software supply chain through three complementary abstractions that characterize what its software \emph{does}, what it \emph{handles}, and what it \emph{runs on}. \emph{Transformations} are the processes the software performs, the code that derives new artifacts from existing ones. \emph{Artifacts} are the persistent objects this software produces, consumes, and exchanges, with model weights the most notable among them. \emph{Infrastructure} is the computational substrate, itself composed of software, on which transformations execute.

\begin{figure*}[t]
\centering
\includegraphics[width=\textwidth]{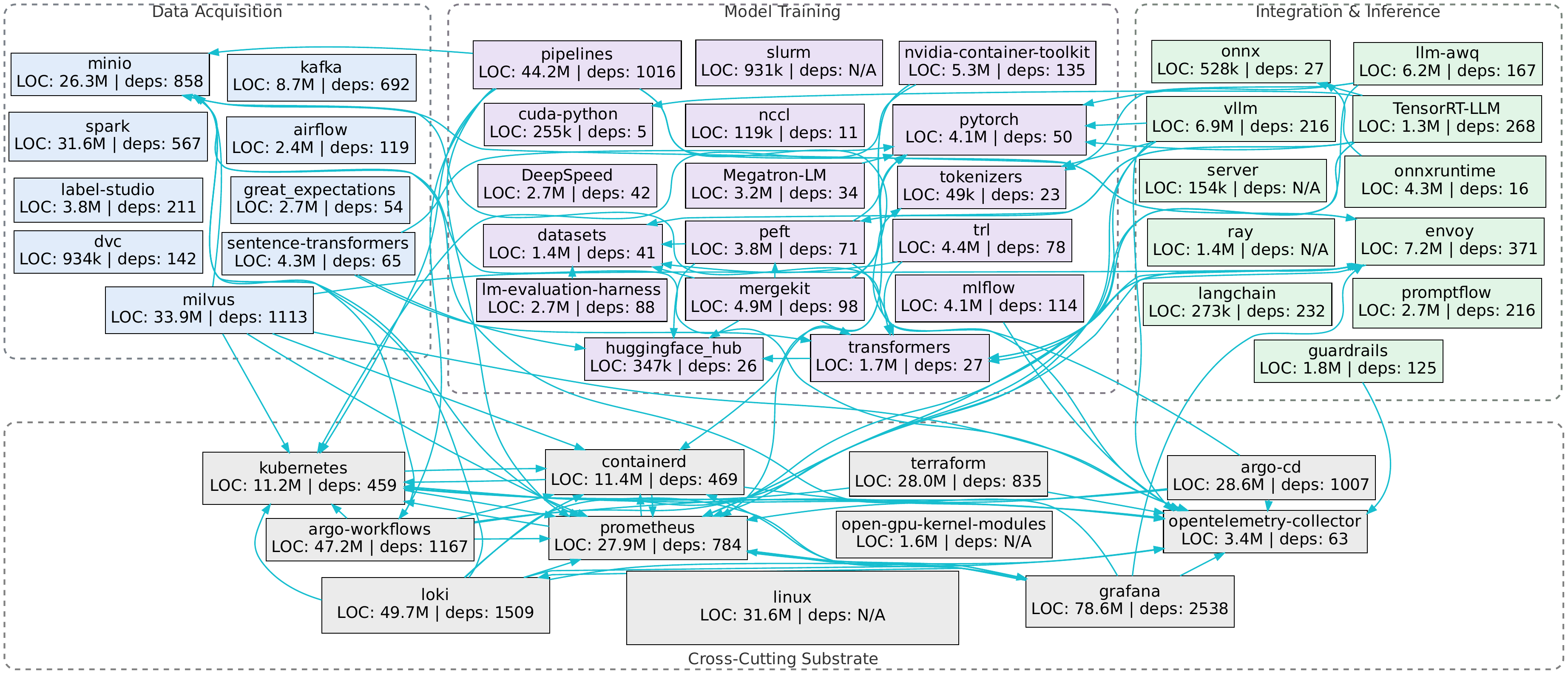}
\caption{The software supply chain of a reference stack of \textbf{48 open-source projects}, combined in an end-to-end AI system. Nodes are software projects; an edge $A \rightarrow B$ indicates that project $A$ uses project $B$. This reference stack totals $\approx 392\,\text{M LOC}$ (see Section~\ref{sec:scale})}.
\label{fig:ai-supply-chain-graph}
\end{figure*}

\begin{definitionbox}
An \emph{AI artifact} is any persistent object produced, consumed, or manipulated software components in the supply chain; the most notable artifact is the model weights.
\end{definitionbox}
\noindent\textit{Example. A released Llama-3 model checkpoint, its tokenizer file, and the preprocessed training corpus used to train it are artifacts handled by the software supply chain.}\\

Every layer's software produces, consumes, or transforms artifacts. For example, data-production software produces datasets and tokenizers. Training software produces model checkpoints and their derivatives. Inference software produces and consumes prompt templates, retrieval indexes, and serving manifests. Cross-cutting infrastructure software produces container images and execution logs. 

The key distinction among artifacts is between \emph{symbolic} and \emph{trained} artifacts. Container images, training scripts, prompt templates, and serving manifests are symbolic artifacts: they specify behavior through explicit human- or machine-written logic; their behavior can be inferred by inspecting their contents or regenerating them from source. Model checkpoints and their derivatives, including fine-tuning adapters, distilled models, merged checkpoints, and quantized variants, are trained artifacts: they encode behavior in learned parameters shaped by data, optimization dynamics, and stochastic processes; their behavior cannot be inferred from inspection alone. This distinction is the source of several open problems examined in Section~\ref{sec:challenges}: the software supply chain can diff, audit, and hash-verify symbolic artifacts, but cannot do the same for trained artifacts, leaving gaps in verification, lineage, and versioning that have no direct analogue in traditional software supply chains.

\newpage
\begin{definitionbox}
An \emph{AI transformation} is a software-implemented process that produces one or more output artifacts from one or more input artifacts.
\end{definitionbox}
\noindent\textit{Example. Fine-tuning a pretrained Llama-3 checkpoint on an instruction dataset produces a new model artifact derived from an existing model artifact and a dataset artifact.}\\

Every architectural layer of an AI system implements transformations. 
Software in the data-production layer implements ingestion, preprocessing, tokenization, chunking, embedding generation, and vector index construction. Training software implements pretraining, fine-tuning, distillation, checkpoint merging, and evaluation. Inference software implements prompt assembly and chaining, compilation, guardrail filtering, quantization, serialization, retrieval, and tool invocation. Cross-cutting infrastructure software implements containerization, log aggregation, and monitoring. Table~\ref{tab:ai-transformations} catalogs these transformations by layer. Transformations also vary in determinism: format conversion and serialization typically produce identical outputs from identical inputs, while training and fine-tuning do not~\cite{pham2020problems}.

\begin{table}[t]
\centering
\footnotesize
\begin{tabularx}{\textwidth}{l l X}
\toprule
\textbf{Layer} & \textbf{Transformation} & \textbf{Description} \\
\midrule

\multirow{7}{*}{\shortstack{Data Acquisition \\ (Section~\ref{sec:data})}}
& \cellcolor{datarow}Data ingestion & \cellcolor{datarow} Importing external data into datasets (e.g. Apache Kafka). \\
& \cellcolor{datarow}Data preprocessing & \cellcolor{datarow}Cleaning and transforming data (e.g. Apache Spark). \\
& \cellcolor{datarow}Data annotation & \cellcolor{datarow}Attaching labels to raw inputs (e.g., Label Studio). \\
& \cellcolor{datarow}Chunking & \cellcolor{datarow}Splitting data into smaller units (e.g. LlamaIndex). \\
& \cellcolor{datarow}Tokenization & \cellcolor{datarow}Converting text into tokens (e.g. HuggingFace Tokenizers). \\
& \cellcolor{datarow}Embedding generation & \cellcolor{datarow}Encoding data as vectors (e.g. sentence-transformers). \\
& \cellcolor{datarow}Vector index construction & \cellcolor{datarow}Building searchable vector indexes (e.g. FAISS). \\

\midrule

\multirow{6}{*}{\shortstack{Training \\ (Section~\ref{sec:training})}}
& \cellcolor{trainingrow}Model training & \cellcolor{trainingrow}Learning model parameters (e.g. PyTorch). \\
& \cellcolor{trainingrow}Fine-tuning & \cellcolor{trainingrow}Adapting pretrained models (e.g. PEFT). \\
& \cellcolor{trainingrow}Distillation & \cellcolor{trainingrow}Compressing models via imitation (e.g. PyTorch). \\
& \cellcolor{trainingrow}Checkpoint merging & \cellcolor{trainingrow}Combining model checkpoints (e.g. mergekit). \\
& \cellcolor{trainingrow}Preference-based alignment & \cellcolor{trainingrow}Tuning model behavior using preferences (e.g., RLHF)\\
& \cellcolor{trainingrow}Evaluation & \cellcolor{trainingrow}Evaluating artifacts before deployment (e.g. HELM). \\

\midrule

\multirow{9}{*}{\shortstack{Integration \\ \& Inference \\  (Section~\ref{sec:integration})}}
& \cellcolor{inferencerow}Prompt engineering & \cellcolor{inferencerow}Creating and refining prompts (e.g. PromptFlow). \\
& \cellcolor{inferencerow}Prompt chaining & \cellcolor{inferencerow}Linking prompts into workflows (e.g. LangChain). \\
& \cellcolor{inferencerow}Compilation & \cellcolor{inferencerow}Converting models to runtime formats (e.g. ONNX). \\
& \cellcolor{inferencerow}Guardrail filtering & \cellcolor{inferencerow}Enforcing input/output policies (e.g. Guardrails AI). \\
& \cellcolor{inferencerow}Quantization & \cellcolor{inferencerow}Reducing precision for efficient inference (e.g. AWQ). \\
& \cellcolor{inferencerow}Serialization & \cellcolor{inferencerow}Packaging models for deployment (e.g. HuggingFace Hub). \\
& \cellcolor{inferencerow}Retrieval & \cellcolor{inferencerow}Fetching context at runtime for inference (e.g. vLLM). \\
& \cellcolor{inferencerow}Tool invocation & \cellcolor{inferencerow}Executing external API calls (e.g. LangChain). \\
& \cellcolor{inferencerow}System serving & \cellcolor{inferencerow}Serving and exposing inference APIs (e.g. Ray Serve). \\

\midrule

\multirow{3}{*}{\shortstack{Cross-cutting \\ (Section~\ref{sec:cross-cutting})}}
& \cellcolor{crossrow}Containerization & \cellcolor{crossrow}Packaging into runtime images (e.g. containerd). \\
& \cellcolor{crossrow}Log aggregation & \cellcolor{crossrow}Collecting execution logs (e.g. Prometheus). \\
& \cellcolor{crossrow}Monitoring and feedback & \cellcolor{crossrow}Using telemetry to adapt systems (e.g. OpenTelemetry). \\

\bottomrule
\end{tabularx}
\caption{Common transformations in AI supply chains, all mediated by software.}
\label{tab:ai-transformations}
\end{table}

\begin{definitionbox}
\emph{AI infrastructure} comprises the computational platforms, storage systems, and operational environments on which AI transformations and processes execute.
\end{definitionbox}
\noindent\textit{Example. An operator fine-tunes a model on a SLURM-managed GPU cluster, stores checkpoints in S3, and serves the model through vLLM behind an Envoy gateway. Each component in this chain is an infrastructure dependency of the resulting system.}\\

Infrastructure software, like transformations, is present in every layer: data lakes and stream processing platforms underpin data pipelines; GPU clusters and experiment trackers underpin training; model registries, vector databases, and serving frameworks underpin inference; logging and monitoring systems underpin the cross-cutting layer. A key distinction is between \emph{internal infrastructure} (self-managed clusters, on-premise storage) that remains under organizational control, and \emph{external infrastructure} (cloud compute, managed ML platforms, third-party APIs) that reduces operational burden but introduces dependencies on external providers. Most production systems rely on a mix, creating supply chains that span trust boundaries.

Some software in the supply chain plays a specific role: supporting the tracking, versioning, and distribution of artifacts rather than transforming them. Dataset management systems, experiment trackers, and model registries are themselves software components within the supply chain, and what they contribute is metadata and indirection. Throughout the paper we are careful to keep this distinction: a registry is a piece of software in the supply chain, while the checkpoint it hosts is an artifact handled by that software.

Taken together, the software components described by these three abstractions constitute the supply chain under study. The primary dependency graph of this supply chain has software components as nodes and `A~uses~B' relations as edges; Figure~\ref{fig:ai-supply-chain-graph} instantiates this graph for a reference stack of 48 open-source projects, and Section~\ref{sec:scale} measures its scale empirically. Artifact derivations form a secondary, analytically useful graph, the \emph{lineage graph}, whose nodes are artifacts and whose edges are transformations. The lineage graph is not itself the supply chain but a property that the supply chain's software is expected to record and expose. As Section~\ref{sec:lineage} argues, current supply chain software records this graph only partially, and that partial representation is one of the four structural gaps this paper diagnoses. The following four sections examine the supply chain layer by layer.

\section{Data Acquisition Pipelines}
\label{sec:data}
Data acquisition pipelines form the data layer of the AI software supply chain (blue components in Figure~\ref{fig:ai-supply-chain-graph}), upstream of training and serving in the dependency graph. They produce and maintain the data artifacts that flow into every downstream layer: \emph{training datasets} consumed by model development, \emph{retrieval corpora} queried by deployed systems at inference time, and \emph{inference logs and feedback traces} captured from production systems and routed back into training, evaluation, and monitoring. The last of these closes the loop between serving and training, making the system lifecycle continuous rather than one-way.

We organize this layer in three stages: \emph{ingestion}, which acquires raw inputs from external sources;  \emph{dataset construction}, which transforms raw inputs into training-ready datasets; and \emph{inference-time data pipelines}, which build the retrieval corpora queried by deployed systems and capture the logs and feedback they produce.

\subsection{Data Sourcing and Ingestion}
\transformation{Data ingestion} is the process by which raw inputs enter the AI supply chain. It introduces two classes of dependencies: the \emph{external data sources} that supply the raw inputs, and the \emph{ingestion pipeline software} that acquires, routes, and persists them.

\paragraph{External data sources.}
Every data source, whether a web crawl, a third-party data feed, or an internal enterprise system, constitutes an \artifact{external data source}: a supply chain dependency whose availability, format, and content may change without notice and without any corresponding version event in the consuming pipeline. A vendor may alter its data cleaning procedures, a source system may migrate and change field semantics, or a crawling service may modify its deduplication logic, and in each case the change propagates silently into the downstream dataset unless the ingestion pipeline explicitly records the source state at acquisition time.

\paragraph{Ingestion pipeline software.}
Ingestion is performed by workflow orchestrators such as \href{https://airflow.apache.org/}{Apache Airflow} and message brokers such as \href{https://kafka.apache.org/}{Apache Kafka}, governed by persistent \artifact{ingestion pipeline definitions}: DAG specifications, topic configurations, retry policies, and data routing rules. These configurations shape which data reaches downstream processes and under what conditions. A change to a routing rule or a retry policy can alter the composition of a training dataset without any modification to the data sources themselves or to the training code that consumes them. The pipeline software is itself a supply chain dependency: orchestrator versions, connector plugins, and message broker clients are pinned in manifests that propagate to every pipeline run.

Together, these two classes of dependency produce \artifact{ingestion artifacts}: raw document collections, source snapshots, and ingestion metadata recording acquisition timestamps, source identifiers, and schema versions. The metadata is what downstream layers can rely on to reconstruct which source state and which pipeline configuration shaped a given dataset, and its completeness determines how far backward lineage can be traced (Section~\ref{sec:lineage}).

\subsection{Dataset Construction and Preprocessing}
Once raw inputs have been ingested, they must be transformed into dataset artifacts that downstream training can consume. Two transformations dominate this step: \emph{annotation}, which attaches labels or preference signals to raw inputs, and \emph{preprocessing}, which cleans and restructures them into a form suitable for training.

\paragraph{Annotation.}
Many training regimes, including supervised fine-tuning and preference-based alignment (Section~\ref{sec:training}), require each input to be paired with a target label, a preference ranking, or a quality score. \transformation{Data annotation} produces these pairings through human annotators, automated labeling systems, or a combination of both, guided by an \artifact{annotation schema} that specifies the label space, instructions, and adjudication policy. Annotation platforms such as \href{https://labelstud.io/}{Label Studio} manage labeling workflows and produce \artifact{labeled dataset artifacts} whose properties depend on the schema version, annotator pool, and adjudication policy.

\paragraph{Preprocessing.}
\transformation{Data preprocessing} pipelines transform raw or labeled inputs into training-ready form, performing filtering, deduplication, normalization, parsing, partitioning, and feature extraction. The resulting \artifact{dataset artifacts} include cleaned and filtered corpora, train/validation/test splits, feature tables, and preprocessing metadata describing provenance, preprocessing parameters, and filtering decisions. \transformation{Tokenization} is a preprocessing step whose output is consumed directly by training, and whose behavior is governed by a \artifact{tokenizer configuration}: a vocabulary file that becomes a shared supply chain dependency of both training and inference.

\paragraph{Dataset versioning.}
Dataset management systems such as \href{https://dvc.org/}{DVC} and \href{https://lakefs.io/}{lakeFS} are examples of software in the supply chain whose role is specifically to support lineage tracking for the artifacts the supply chain produces. Each commit or tag in these systems produces a \artifact{data version snapshot}: an immutable reference that pins a dataset to a specific state, allowing downstream training runs to declare a dependency on a reproducible data configuration even while upstream ingestion and preprocessing continue to evolve.

\subsection{Inference-Time Data Pipelines}
Alongside the pipeline for producing training datasets, two further classes of data pipeline operate at the boundary with deployed systems: \emph{retrieval-corpus pipelines}, which build and maintain the document stores that inference queries consume, and \emph{inference-capture pipelines}, which record inference events and feedback for later reuse. 

\paragraph{Retrieval corpora.}
A \artifact{retrieval corpus} is a collection of documents, embedding vectors, and indexes queried by a deployed system at inference time to supply external context (Section~\ref{sec:integration} describes its integration into serving). Unlike training datasets, which shape model parameters once and then remain fixed at the version pinned by the training run, retrieval corpora supply context to a running system and can be updated independently of the model itself. Building a retrieval corpus chains three tightly coupled transformations: \transformation{chunking} splits source documents into passages, \transformation{embedding generation} encodes each passage as a dense vector, and \transformation{vector index construction} organizes the vectors into a searchable index (e.g., \href{https://milvus.io/}{Milvus}, FAISS~\cite{johnson2019billion}). Retrieval corpora are often updated continuously through periodic refresh or streaming ingestion to reflect new documents or changes in source systems.

\paragraph{Inference capture.}
Inference-capture pipelines record the activity of deployed systems, producing \artifact{inference logs} (requests, model outputs, retrieval results, tool invocations, latency and error metadata) and \artifact{feedback traces} (explicit user ratings, implicit signals such as edit distance or session abandonment, and downstream outcome labels). These artifacts are routed back into the data layer through several channels: training-data pipelines consume them as inputs for supervised fine-tuning, and evaluation pipelines use them to construct regression suites grounded in production distributions.

\section{Model Training Infrastructure}
\label{sec:training}

The training layer (purple components in Figure~\ref{fig:ai-supply-chain-graph}) consumes dataset artifacts and produces model checkpoints whose properties depend on both the upstream data and the infrastructure that executes training. We organize this layer around three concerns: \emph{training orchestration and execution}, which coordinates training steps and provides the hardware and runtime stack that execute them; \emph{training variants}, which enumerates the principal procedures (pretraining, fine-tuning, distillation, merging) by which checkpoints are produced from upstream artifacts; and \emph{provenance and distribution}, which tracks how a checkpoint was produced and how it reaches downstream consumers.

\subsection{Training Orchestration and Environment}
Training workflows coordinate a sequence of dependent steps: dataset loading, model initialization, iterative optimization, periodic evaluation, and artifact export. Orchestration frameworks such as \href{https://www.kubeflow.org/docs/components/pipelines/}{Kubeflow Pipelines} or \href{https://airflow.apache.org/}{Apache Airflow} express these steps as directed pipelines, managing job dependencies, hyperparameter configurations, and scheduling policies. The pipeline structure is encoded in a persistent \artifact{training-pipeline definition} that declares the dependency order among steps, the hyperparameter search space, and the conditions under which steps are skipped or retried. The orchestration layer also governs the lifecycle of intermediate artifacts, determining when checkpoints are saved, when evaluation runs are triggered, and when results are exported.

Training environments are typically packaged as container images (Section~\ref{sec:cross-cutting}). Within these environments, the training logic is implemented using machine learning frameworks such as \href{https://pytorch.org/}{PyTorch} or \href{https://github.com/google/jax}{JAX}. Large-scale training jobs frequently span multiple GPUs or compute nodes, relying on communication libraries such as \href{https://github.com/NVIDIA/nccl}{NVIDIA NCCL} for gradient synchronization and hardware runtimes such as \href{https://developer.nvidia.com/cuda-toolkit}{CUDA} for accelerated execution. Each of these components, including framework version, communication library, GPU driver, container base image, is a software dependency of the supply chain that produces the resulting checkpoint.

\subsection{Multi-Model Training}
Checkpoints are produced through several distinct procedures, each introducing a different pattern of supply chain dependencies on upstream artifacts. 
\transformation{Model training} (pretraining) learns parameters from scratch on a large dataset, producing a \artifact{base checkpoint} whose dependencies are a training dataset, a training-pipeline definition, and the execution environment described above; it is the most resource-intensive variant and is typically performed only by a small number of organizations, so downstream actors consume pretrained checkpoints as starting points for the variants below. 
\transformation{Fine-tuning} adapts a pretrained checkpoint to a specific task or behavior using a smaller, task-specific dataset, either through full-parameter updates that subsume the base model or through parameter-efficient methods such as LoRA~\cite{hu2022lora}, which freeze the base checkpoint and train a small \artifact{adapter artifact} separately. The serving software must load the adapter against the exact base checkpoint it was trained against; loading it onto a different base version produces undefined behavior. 
\transformation{Preference-based alignment} tunes model behavior using datasets of ranked preferences rather than labeled outputs, through methods such as RLHF or direct preference optimization, and inherits dependencies from both a base checkpoint and a preference dataset whose construction is itself a supply-chain concern. 
\transformation{Distillation} produces a smaller \artifact{student checkpoint} by training it to imitate the outputs of a larger \artifact{teacher checkpoint} on a distillation dataset, and the student therefore depends on both parents directly. 
\transformation{Checkpoint merging} combines parameters from multiple trained checkpoints into a single artifact through weighted averaging or task-specific interpolation, without re-executing training, so the merged checkpoint inherits from every source checkpoint independently.

\subsection{Provenance and Model Registries}
Training pipelines continuously produce intermediate artifacts, including optimizer states, evaluation metrics, and training logs, that must be stored, versioned, and referenced by downstream processes. Experiment tracking systems such as \href{https://mlflow.org/}{MLflow} and \href{https://wandb.ai/}{Weights \& Biases} aggregate these outputs together with run configuration (hyperparameters, dataset references, infrastructure settings) into a persistent \artifact{experiment metadata record}.

Experiment tracking software is the principal software component supporting \emph{provenance} for trained checkpoints. The experiment metadata record it produces is the recorded history of the inputs, transformations, and environment from which the checkpoint was derived. Unlike the build logs of compiled software, which can in principle be regenerated by re-executing the build, training provenance cannot be reconstructed after the fact, non-determinism in training (Section~\ref{sec:verifiability}) makes the record the only authoritative account of how a checkpoint came to be.

Artifacts are promoted into model registries such as \href{https://huggingface.co/}{HuggingFace Hub} or \href{https://mlflow.org/docs/latest/ml/model-registry}{MLflow Model Registry}, which manage versioning, lifecycle transitions, and deployment readiness. Registries are the software through which training-layer artifacts reach every downstream consumer. A \artifact{model checkpoint} and its associated \artifact{tokenizer} are typically versioned and distributed together, because their vocabularies must remain aligned for inputs to round-trip correctly. Registries also host the adapter, student, and merged checkpoints produced by the variants above, making them the junction point at which the diverse derivation structures of the training layer converge into a shared distribution surface.

\section{Inference Infrastructure \& Final Integration}
\label{sec:integration}
At this layer (green components in Figure~\ref{fig:ai-supply-chain-graph}), orchestration software composes trained checkpoints with the auxiliary components that turn a model into a system: retrieval pipelines that fetch external context, tool adapters that invoke external services, prompt templates that structure model inputs, and guardrails that constrain model outputs. The resulting assembly is then deployed within runtime infrastructure that executes inference at scale.

In the following, we organize this layer around four concerns.

\subsection{Evaluation and Release Gating}

\transformation{Evaluation} and release gating software sits at the boundary between training and deployment, controlling which artifact versions flow downstream into production. Evaluation frameworks such as \href{https://crfm.stanford.edu/helm/}{HELM} or \href{https://github.com/EleutherAI/lm-evaluation-harness}{lm-eval-harness} execute benchmark suites, adversarial prompts, and red-teaming tests against candidate checkpoints and system configurations. The resulting \artifact{evaluation artifacts}, which include benchmark scores, regression test reports, and safety audit records, document the evidence on which promotion decisions rest. The release gate's configuration, which benchmarks to run, what thresholds to enforce, which rollout policy to follow, is itself a configuration artifact within the software supply chain.

\subsection{Application Orchestration}
Modern compound AI systems rarely invoke a single model in isolation. Instead, orchestration code coordinates sequences of model calls alongside auxiliary components such as retrieval steps, verification passes, tool invocations, and conditional routing. This layer introduces \artifact{orchestration configurations} into the supply chain: prompt templates that define the structured inputs controlling model behavior at each step (\transformation{prompt engineering}), routing policies that determine which model or serving endpoint handles a given request, and sampling parameters (e.g., temperature, top-k) that govern generation behavior.

Two orchestration patterns dominate in practice, both commonly realized through frameworks such as \href{https://www.langchain.com/}{LangChain} and \href{https://www.llamaindex.ai/}{LlamaIndex}. \transformation{Retrieval-augmented generation} (RAG) retrieves documents from an external corpus and appends them to the model's input before generation, making the \artifact{retrieval corpus}, the embedding model, and the retrieval policy into runtime artifact dependencies of the orchestration software serving every response (Section~\ref{sec:data} describes how these upstream artifacts are constructed and maintained). \transformation{Prompt chaining} makes the dataflow between prompts explicit: a \artifact{prompt chain definition} specifies which prompts feed which downstream steps and what output format each must produce~\cite{clariso2023model}.

\subsection{Tool Integration and Guardrails}
The assembled pipeline depends not only on models and orchestration code but also on external services and policy components wired in during integration. These extend the dependency graph in two directions: outward, through tool integrations that expose the system to external APIs, and inward, through guardrail mechanisms that constrain what the system can produce.

\paragraph{Tool integration.}
In the typical tool-use pattern, a model generates a structured request, a \transformation{tool invocation} (via a tool adapter) translates it into an API call, and the result is returned for further reasoning. Each integration is governed by a \artifact{tool schema} (e.g., an \href{https://platform.openai.com/docs/guides/function-calling}{OpenAI function-calling} definition or an \href{https://modelcontextprotocol.io/}{MCP} tool manifest) that declares input parameters, output format, and invocation semantics. Tool schemas are artifact dependencies of the orchestration software that invokes them.

\paragraph{Guardrails.}
\transformation{Guardrail mechanisms} (e.g., \href{https://guardrailsai.com/}{Guardrails.ai}, \href{https://lmql.ai/}{LMQL}) function as transformations within the pipeline: they receive model outputs and produce filtered or modified outputs. Because guardrails may block, rewrite, or trigger retries of model responses, the observable behavior of a deployed system can differ substantially from the behavior of the underlying model evaluated in isolation. \artifact{Guardrail configurations}, rule sets, content filter lists, and output schema definitions, are safety-critical configuration artifacts consumed by guardrail software across every pipeline that references them.

\subsection{Inference Execution}
Before an assembled pipeline can serve requests, its model artifacts must be adapted to the serving environment, and the infrastructure that executes inference must be configured.

\paragraph{Runtime model transformations.}
Model artifacts frequently undergo runtime transformations that produce a derived artifact under the same nominal identity (e.g., `Llama-3 70B checkpoint') but with different numerical behavior. 
The most consequential is \transformation{quantization}: reducing the numerical precision of model parameters to decrease memory footprint and increase inference throughput, which can measurably affect model accuracy on tasks that require precise numerical reasoning or operate near decision boundaries~\cite{NEURIPS2022_c3ba4962}. 
Others include \transformation{compilation} into runtime-specific intermediate representations (e.g., \href{https://github.com/NVIDIA/TensorRT}{TensorRT} engine plans, \href{https://onnx.ai/}{ONNX} files) and \transformation{serialization} (converting model formats across frameworks). 

\paragraph{Serving infrastructure.}
The \transformation{serving} stack introduces additional configuration artifacts and software dependencies. \artifact{Serving configurations}, including deployment manifests, version routing rules, autoscaling policies, and inference engine settings, determine how a trained model behaves in production independent of the model artifact itself. The serving stack introduces three control points where configuration, rather than any artifact change, determines system behavior. At the network boundary, API gateway routing configurations (e.g., \href{https://www.envoyproxy.io/}{Envoy}) control which clients receive outputs from which model version. Behind the gateway, deployment manifests and traffic-splitting rules (managed by frameworks such as \href{https://kserve.github.io/website/}{KServe} or \href{https://docs.ray.io/en/latest/serve/index.html}{Ray Serve}) can redirect production traffic to a different model version without any change to the model artifact itself. At the execution layer, inference engines such as \href{https://github.com/vllm-project/vllm}{vLLM} translate model operations into hardware-specific kernels whose implementation choices can alter numerical results even when the upstream checkpoint is unchanged.

\section{Cross-Cutting Substrate}
\label{sec:cross-cutting}
The three supply chain layers described above do not execute in isolation but share a common operational substrate (grey components in Figure~\ref{fig:ai-supply-chain-graph}), whose components appear as transitive dependencies in every layer. This substrate is a high-leverage point in the AI supply chain: compromising a shared infrastructure component can impact data pipelines, training, and serving simultaneously (as observed in LiteLLM, Section~\ref{sec:intro}), without targeting AI-specific components.

\paragraph{Container orchestration and base images.}
\transformation{Containerization} packages workloads into \artifact{container images} that bundle application code with runtime dependencies (ML frameworks, system libraries, CUDA toolkits) onto clusters managed by \href{https://kubernetes.io/}{Kubernetes} or \href{https://slurm.schedmd.com/}{Slurm}. Prebuilt images from registries such as \href{https://catalog.ngc.nvidia.com/}{NVIDIA NGC} often serve as shared base environments for both training and serving. A shared base image uses transitive dependencies that propagate to every derived workload: when NVIDIA patches an NGC image, the update simultaneously changes the execution environment of training jobs and serving deployments that inherit from it. Container images are also opaque compositions: a \texttt{Dockerfile} may chain multiple upstream images and install packages from several ecosystems, producing a final image whose effective set of supply chain dependencies is the union of all layers but is declared in no single manifest.

\paragraph{Build, deploy, and provisioning systems.}
CI/CD pipelines and GitOps controllers (e.g., \href{https://argo-cd.readthedocs.io/}{Argo CD}, \href{https://argoproj.github.io/workflows/}{Argo Workflows}) hold credentials to package registries, model registries, container registries, and production clusters. A compromised pipeline can inject malicious code into build output it produces while emitting valid signatures, because the artifact was produced by the legitimate build system. This is what occurred in the LiteLLM compromise. For AI supply chains, the blast radius is amplified because a single CI/CD system may build container images, publish Python packages, export model checkpoints, and deploy serving configurations, spanning all four layers. Infrastructure-as-code definitions (e.g., \href{https://www.terraform.io/}{Terraform}) introduce a related risk at the provisioning layer: they determine hardware topology, network isolation, and access control.

\paragraph{Observability stack and runtime.}
\transformation{Log aggregation} (e.g., \href{https://prometheus.io/}{Prometheus}, \href{https://grafana.com/oss/loki/}{Grafana Loki}) and \transformation{monitoring and feedback} systems (e.g., \href{https://opentelemetry.io/}{OpenTelemetry}) are themselves supply chain dependencies that define what is observable. At the base of the stack, the Linux kernel and GPU drivers influence numerical behavior across every layer. A driver update that changes the implementation of a floating-point kernel can alter training convergence and inference outputs simultaneously.

\section{Open Challenges}
\label{sec:challenges}

The preceding sections examined the AI supply chain layer by layer. This section identifies four important properties that the current software supply chain of AI systems fails to provide: 
\emph{verifiability} is about whether an AI artifact is what its producer claims it to be; 
\emph{versioning} is about whether the components of an AI system can be pinned into a coherent, reproducible assembly; 
\emph{observability} relates to whether changes in runtime behavior are detectable and attributable to a specific upstream artifact; and 
\emph{traceability} is the property of deployed AI artifacts to have their dependency graph constructed backward to all its inputs.
Each subsection defines the property formally, characterizes the gap in today's infrastructure, and identifies promising solutions.

\subsection{The Verifiability Gap}
\label{sec:verifiability}

\emph{Verifiability} is the property that any party admitting an AI artifact into a trust perimeter can independently establish that the artifact is what its producer claims it to be. When verifiability holds, tampering with AI artifacts is detectable without trust in the producer or in the infrastructure that transported or stored the artifact.

In traditional software, reproducible builds close a class of attacks in which a compromised build environment injects malicious code: given the same source and build environment, the build process should produce a bit-identical output~\cite{lamb2021reproducible}. An independent party can then recompile the artifact, and a hash mismatch reveals tampering.

The software supply chain of AI systems does not meet this property for trained artifacts.
Training is inherently non-reproducible: GPU scheduling order, floating-point non-associativity, and framework-internal operator selection introduce variation that compounds over millions of gradient steps~\cite{pham2020problems}; two runs of the same procedure on the same infrastructure produce different weights. A checkpoint's hash authenticates the file but reveals nothing about the process that produced it. The problem is not confined to the core training loop. The execution environment beneath (kernel version, driver stack, CUDA runtime, base container image) can alter numerical behavior of both training and inference. 

The consequence for admission-time verification is direct: no independent party can recompute or verify an AI model and detect tampering by hash mismatch. Consider an attacker who compromises a legitimate training cluster and produces a backdoored checkpoint. The checkpoint still carries a valid attestation chain. A re-execution on clean infrastructure would not produce a bit-identical result, so a hash mismatch proves nothing. SLSA~\cite{SLSA} and in-toto~\cite{torres2019toto} do not close this gap; they relocate it. The trust anchor shifts from the artifact to the attestor, and an adversary who compromises attestation infrastructure, or the training environment it signs off on, can produce valid provenance for a poisoned training run.

\textit{Directions.}
Because no canonical output exists to check against, output-equality verification is ruled out, and admission-time verification of trained artifacts must instead seek process-level assurance: attesting that a specific, auditable procedure was followed under controlled conditions. The most tractable direction is \emph{compositional verification}: decomposing the pipeline into deterministic segments (e.g., preprocessing, serialization, quantization) that admit conventional hash-based verification, and confining process-level attestation to the irreducibly stochastic training step. This shrinks the unverifiable surface, and reduces the problem to a small residual that is explicit and bounded.
Two other directions are worth noting: \emph{bounded non-determinism}, characterizing the expected output distribution of a training procedure so that re-executions can be checked for statistical rather than bit-level consistency~\cite{semmelrock2025reproducibility}; and \emph{hardware-rooted attestation}, using trusted execution environments to anchor attestation in the execution hardware itself~\cite{dhanuskodi2023creating,vaswani2023confidential}.
Both are promising but do not yet exist at the scale of frontier training runs.

\subsection{The Versioning Gap}
\label{sec:versioning}

\emph{Versioning} is the property that a software system can be pinned to a coherent, reproducible assembly of artifact versions whose behavioral compatibility is declared and enforced. When it holds, an operator can identify, reconstruct, and roll back any past deployed configuration of the system.

Package managers enforce immutability, version constraints, lockfiles pin transitive dependencies, and build systems ensure that a declared dependency graph matches the artifacts actually consumed. These mechanisms work because an application-level dependency graph is explicit and version identifiers are stable references to immutable objects.

The software supply chain of AI systems violates both halves of the versioning property: first, a compound AI system has many components with no explicit dependencies; second the \emph{behavioral coupling} between those components is not exposed in any manifest. A LoRA adapter requires the exact base checkpoint it was trained against; a retrieval index is valid only against the embedding model that produced its vectors; a prompt template encodes undeclared assumptions about the format and structure of the upstream model's outputs. None of these couplings are expressed in any dependency specification, and none are enforced by any resolver. An operator who updates the embedding model continues to serve a stale index without obvious error; a model whose output format shifts silently breaks every downstream consumer whose version identifiers remain unchanged.

The gap is starkest at third-party API boundaries. Consider an operator building a production application on a model API. Prompt templates are tuned to the model's specific behavior. Evaluation thresholds are calibrated against its output distribution. Guardrail policies are configured around its refusal patterns. When the provider silently updates the underlying checkpoint~\cite{chen2024chatgpt}, the operator's full artifact assembly, designed and validated as an integrated unit, is suddenly coupled to a model it was never tested against. The available remediation is asymmetric: the operator can roll back their own components but cannot revert the upstream model, because no version selector exists and the previous checkpoint is no longer served. When the provider later deprecates the model family entirely, reproducing the previous system state would require the specific prompt--model--guardrail combination that was validated together, and no lockfile ever captured that joint configuration.

\textit{Directions.}
AI supply chains need the equivalent of a lockfile: a declarative specification that pins the full artifact assembly of a compound AI system to a reproducible, deployable state. Such a specification must handle artifact types pinnable as immutable hashed blobs or as point-in-time snapshots. Such a specification declares the interface contracts (which adapter with which base, which index with which embedding model, which prompt with which output format) so that a resolver can detect a violation. It must also overcome \emph{semantic opacity}: model checkpoints have no version identifier to signal whether an update is backward-compatible on a given input distribution~\cite{chen2024chatgpt}.

The hardest case is \emph{cross-boundary pinning}. A lockfile encodes the consumer's intent to use a specific version, but enforcement requires either a neutral registry that cannot rewrite history, or a local copy of the pinned artifact. Neither condition holds when the upstream is a first-party model provider: versioning is controlled unilaterally, and the weights are typically not available for local caching. Closing this gap requires negotiated versioning contracts, with cryptographically bound snapshots of the served artifact state.

\subsection{The Observability Gap}
\label{sec:observability}

\emph{Observability} is the property that changes in the runtime behavior of a deployed system are detectable from its telemetry, and attributable to the specific upstream artifact whose version changed. When it holds, silent regressions surface as observable signals rather than as accumulated user complaints.

Traditional software supply chains fail loudly. A mismatched dependency produces a compilation error or a runtime exception; a broken interface throws a stack trace that localizes the failure. AI supply chains fail silently. A retrieval corpus re-indexed with a different embedding model returns less relevant context. A guardrail update filters valid responses. A quantized checkpoint produces subtly different outputs on precision-sensitive inputs. In each case the system continues to run and returns plausible results. No error is raised, no exception propagates, and a latency dashboard reporting healthy p99 response times will not surface any behavioral changes. Behavioral drift must instead be inferred from the statistical properties of responses over time, and current telemetry is not designed for this.

Consider a production RAG system that begins returning lower-quality responses. During the same two-week window, three upstream artifacts changed independently: the retrieval corpus was re-indexed after an embedding model update, a guardrail was tightened, and the system prompt was edited. Operational dashboards show healthy latency and error rates within normal bounds.
Without per-request metadata recording which artifact versions produced each response, even detecting that something has regressed requires a behavioral baseline that current serving stacks do not maintain. The team may notice only when user complaints accumulate, by which point the change window has widened further and the signal is buried in noise.

\textit{Directions.}
AI supply chain observability requires two capabilities that current AI infrastructure does not provide. 
The first is \emph{artifact-aware instrumentation}: every response must be tagged, at serving time, with the identifiers of the checkpoint, prompt template, retrieval index snapshot, guardrail configuration, and decoding parameters that produced it, at a granularity sufficient to distinguish a changed corpus from an unchanged one and sustainable at production request volumes. This is a schema problem as much as a storage problem: the tags must be machine-comparable across versions, which in turn requires versioning the upstream artifacts in the sense of Section~\ref{sec:versioning}.
The second is \emph{behavioral drift detection}: the stream of artifact-annotated responses must be monitored for statistical shifts in response distributions (semantic embeddings of outputs, refusal rates, length distributions, retrieval-hit patterns), with any shift localized in time and correlated against the artifact-change log. Neither capability exists in mainstream serving stacks today.

\subsection{The Traceability Gap}
\label{sec:lineage}

\emph{Traceability} is the property that for any artifact in the system, the full set of its ancestors and the full set of its descendants can be enumerated and verified. When it holds, the impact of a compromised input can be bounded forward to every affected deployment, and the origin of an observed defect can be traced backward to its source.

Software supply chains rely on build-graph \emph{lineage} to track how each binary derives from its inputs (source files, compiler version, build flags, dependency versions).
The build graph is tractable for two reasons. Transformations are deterministic, so any party can recompute an artifact from its declared inputs. Derivations are single-parent, so the graph is a tree whose edges are cheap to enumerate and verify.

In AI systems, lineage must capture not only \emph{what} went into each artifact (datasets, configurations, base checkpoints) but \emph{how} the non-deterministic transformation executed (learning rate schedules, optimizer state, hardware topology), and it must include \emph{policy decisions}: the safety thresholds, benchmark gates, and promotion criteria that determined which candidate checkpoint reached production. Policy decisions are first-class nodes in the lineage graph, because changing a threshold can redirect which artifact ships without changing any training input or hyperparameter.

The graph is further complicated in both directions.
Backward, multi-parent derivations are common: a distilled checkpoint inherits from both its teacher and its distillation dataset; a merged checkpoint inherits independently from every source (Section~\ref{sec:training}). Tracing an anomaly backward requires traversing every parent lineage.
Forward, runtime transformations (quantization, compilation, serialization) produce derived serving artifacts distinct from the original checkpoint, yet downstream consumers often reference these derivatives under the same nominal identity.
Current experiment trackers, registries, and serving stacks capture this graph only partially.

The cost of fragmentation is concrete. In December 2023, the Stanford Internet Observatory identified over one thousand validated instances of child sexual abuse material in LAION-5B, a dataset of five billion image-text pairs used to train Stable Diffusion and other widely deployed image generation models~\cite{thiel2023identifying}. The discovery triggered an immediate forward-lineage crisis: the Stanford report recommended that models trained on LAION-5B be deprecated and their distribution ceased where feasible. This was impossible to automate because no software connected the dataset to its downstream derivatives.

\textit{Directions.}
Closing the AI traceability gap does not require full transparency, but it requires new software components in the supply chain: specifically, mechanisms for exchanging \emph{lineage attestations}, signed machine-verifiable claims about an artifact's upstream dependencies that convey supply chain properties without revealing proprietary details. Two prerequisites are missing.
First, a \emph{representation}: existing attestation frameworks such as in-toto~\cite{torres2019toto} and SLSA~\cite{SLSA} model single-parent, deterministic transformations, whereas AI lineage requires representing non-deterministic edges, multi-input transformations (checkpoint merging, distillation), policy decisions as first-class nodes, and inference-time artifact assembly. Model cards~\cite{mitchell2019model}, dataset cards~\cite{gebru2021datasheets}, and proposed AI Bills of Materials~\cite{nocera2025we} document individual artifacts but do not connect them into the cross-artifact and cross-layer dependency graph that lineage analysis requires.
Second, a \emph{protocol}: a convention on what each party in a multi-organization supply chain attests, at what granularity, and under what access controls.
The difficulty here is not purely technical: providers have commercial and legal reasons to keep lineage opaque. Training data may be licensed under terms that preclude disclosure, training procedures may constitute trade secrets, and voluntary disclosure may expand liability exposure. A workable protocol must therefore be not only expressive but \emph{incentive-compatible}, likely through minimal-disclosure attestation primitives and, ultimately, regulatory or contractual pressure to make disclosure non-optional for specific properties.

\section{Scale and Shape of the Supply Chain}
\label{sec:scale}

To ground our analysis of the AI supply chain, we define a representative AI stack and measure its software supply chain.

\begin{figure}[t]
  \centering
  \includegraphics[width=\linewidth]{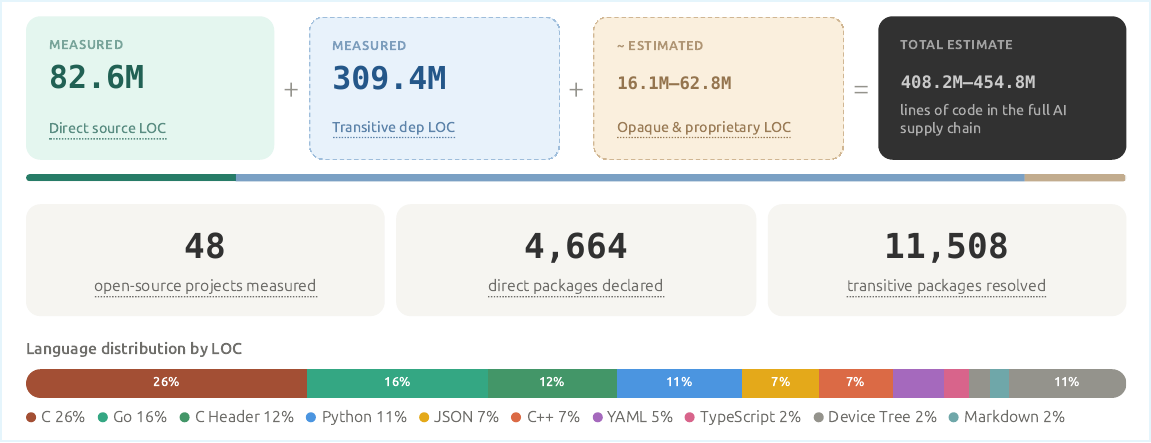}
  \caption{Excerpt of the companion dashboard summarizing the measured AI supply chain. The lower bar shows the language distribution.}
  \label{fig:dashboard}
\end{figure}

We construct the reference stack in three steps. First, for each functional role identified in Sections~\ref{sec:data}--\ref{sec:cross-cutting} (e.g., training orchestrator, vector store, model registry), we enumerate candidate open-source projects, assisted by a large language model prompted to propose projects filling each role with supporting references.
Second, we retain a candidate only if its production use is independently verifiable through a primary source (e.g., an AI lab engineering blog, system paper, etc). Third, we verify that every role is filled by at least one retained project and pin each to a released version for reproducibility. The resulting stack comprises 48 open-source projects spanning the four architectural layers and is representative rather than exhaustive: every functional role is filled by at least one project with documented production use. The companion dashboard\footnote{https://assert-kth.github.io/ai-supply-chain/} lists each project, its pinned version, and the primary source grounding its selection. Figure~\ref{fig:dashboard} shows an excerpt of the dashboard.

For each project in the stack, we construct its dependency graph by resolving both direct and transitive dependencies through the native package manager of the component (PyPI, Go modules, Cargo, Maven, Gradle, npm). For C/C++ components, direct dependencies are extracted statically from build manifests (\texttt{configure.ac}, \texttt{CMakeLists.txt}) and transitive resolution is not attempted. We then measure source lines of code (LOC) with \texttt{scc}, excluding comments, blanks, and generated or vendored files; manifest-detection additionally skips \texttt{test}, \texttt{examples}, \texttt{bench}, and \texttt{vendor} subtrees when enumerating dependencies. 

Figure~\ref{fig:ai-supply-chain-graph} renders the resulting inter-project graph: nodes are projects, colors identify the architectural layer, and an edge $A \rightarrow B$ indicates that $A$ declares a dependency on at least one package of $B$. Edges therefore reflect what package managers can resolve; implicit runtime dependencies (e.g., containers that invoke Linux syscalls, or workloads that rely on GPU drivers loaded at execution time) are not represented. 
The 48 direct projects declare 4{,}664 manifest-listed packages across six ecosystems (PyPI, Go modules, Cargo, Maven, Gradle, npm) and contain 82.6M lines of code, spanning languages including Python, C, C++, Go, Java, Scala, and Rust. Their transitive closure resolves to 11{,}508 packages and an additional 309M lines of code.

Two structural properties stand out. First, the cross-cutting substrate dominates the stack: it alone contributes 51M of the 82.6M direct LOC and the majority of transitive dependencies, so a compromise in a single substrate component reaches artifacts across every layer. The LiteLLM incident (Section~\ref{sec:intro}) is one realization of this: a compromised dependency in the CI/CD substrate propagated across layers because the build infrastructure was a shared upstream node. Second, the bulk of the code an operator implicitly trusts is code they did not select: for every line of code directly chosen, nearly four more enter the stack through transitive resolution.

\section{Related Work}
\label{sec:related}
This paper connects three streams of prior work, each of which addresses a fragment of the AI supply chain without modeling the full dependency structure.

\paragraph{Software Supply Chain Security.}
Traditional software supply chain security has matured around a recurring threat model: malicious code injected through compromised dependencies, build systems, or distribution channels. Ladisa et al.~\cite{ladisa2023sok} provide an attack-tree taxonomy covering 107 vectors mapped to 33 mitigations across open-source ecosystems, and Williams et al.~\cite{williams2025research} synthesize a research agenda spanning component vulnerabilities, build infrastructure compromise, and developer-targeted attacks. The defensive response centers on cryptographic provenance: in-toto~\cite{torres2019toto} produces signed link metadata for each step in a build pipeline, and SLSA~\cite{SLSA} layers progressive assurance levels on top. 
None of this previous work has studied software supply chain security in the context of end-to-end AI systems, as we have done in this paper.

\paragraph{Security and Provenance of ML Artifacts.}
A separate line of work has examined attacks that exploit AI-specific stages of the supply chain. Gu et al.~\cite{gu2017badnets} demonstrate that a maliciously trained checkpoint can pass standard validation while controllable by attacker-chosen inputs, motivating the study of the ML model supply chain as attack surface. Carlini et al.~\cite{carlini2024poisoning} show that web-scale dataset poisoning is practical at $\$60$ cost, and Carlini et al.~\cite{carlini2021poisoning} demonstrate that contaminating $0.01\%$ of a contrastive training corpus suffices to install a backdoor. At the distribution layer, empirical studies of model hubs document a parallel set of risks: Jiang et al.~\cite{jiang2023empirical} report that provenance metadata on Hugging Face is widely incomplete and rarely cryptographically signed, and Siddiq et al.~\cite{siddiq2026empirical} find that remote code execution paths are pervasive in model loading. These studies establish that the AI supply chain is under active attack at the data and model layers; our work situates such findings within a layered architectural decomposition and identifies the structural reasons (non-determinism, behavioral coupling, multi-parent lineage) that make conventional supply chain defenses insufficient against them.

\paragraph{ML Systems Engineering and Documentation.}
At the systems-engineering level, Sculley et al.~\cite{sculley2015hidden} observe that ML systems accumulate hidden technical debt through entanglement and undeclared consumers, explicitly identifying  reproducibility and configuration as recurring sources of operational fragility. To support transparency at the artifact level, Mitchell et al.~\cite{mitchell2019model} propose model cards that document a model's intended use, performance, and limitations, and Gebru et al.~\cite{gebru2021datasheets} propose datasheets describing dataset composition, collection, and known biases. Industry efforts toward AI Bills of Materials extend this idea to a structured inventory of components~\cite{nocera2025we}. These artifacts document individual nodes of the lineage graph but do not connect them: there is no protocol for cross-organization attestation, no mechanism for enforcing declared couplings at deployment time, and no representation for non-deterministic or multi-parent derivations. Our work is much more ambitious in scope and calls for treating the AI supply chain as a whole, not only its individual artifacts.

\section{Conclusion}
\label{sec:conclusion}
This paper is the first to map the software supply chain of AI systems end-to-end. 
The picture that emerges is one of substantial complexity: a compound AI system draws on four architectural layers and three classes of artifacts, composed through pipelines whose dependencies cross several organizational boundaries. 
Our reference stack of 48 open-source projects makes this scale concrete, with 4,664 direct and 11,508 transitive dependencies spanning 392M lines of code.

We have identified four obstacles to a trustworthy AI supply chain: verifiability, versioning, observability, and traceability.
They all stem from a common root: trained artifacts encode behavior through stochastic processes that resist the deterministic assumptions on which existing software supply chain integrity mechanisms depend. Solving these problems requires new research on process-level assurance, heterogeneous dependency coordination, artifact-aware instrumentation, and cross-boundary lineage attestation. Until such mechanisms exist, AI systems will remain vulnerable, offering broad attack surface for supply chain compromise. 
We believe there is no AI safety without AI supply chain integrity.

\section*{Acknowledgements}
This work was supported by the CHAINS Project through Swedish Foundation for Strategic Research (SSF).

\bibliographystyle{plain}
\bibliography{bibliography}

\end{document}